\documentclass[aps,pra,amsmath,amssymb,preprint,letterpaper,superscriptaddress,11pt]{revtex4-1}

\usepackage{hyperref}
\usepackage{graphicx}
\usepackage{dcolumn}

\newcommand{\trans}{\mathsf{T}}

\begin{document}

\title{Excited electronic states from a variational approach based on
  symmetry-projected Hartree--Fock configurations}

\author{Carlos A. Jim\'enez-Hoyos}
\affiliation{Department of Chemistry, Rice University, Houston, TX
  77005}

\author{R. Rodr\'iguez-Guzm\'an}
\affiliation{Department of Chemistry, Rice University, Houston, TX
  77005}
\affiliation{Department of Physics and Astronomy, Rice University,
  Houston, TX 77005}

\author{Gustavo E. Scuseria}
\affiliation{Department of Chemistry, Rice University, Houston, TX
  77005}
\affiliation{Department of Physics and Astronomy, Rice University,
  Houston, TX 77005}

\date{\today}

\begin{abstract}
Recent work from our research group has demonstrated that
symmetry-projected Hartree--Fock (HF) methods provide a compact
representation of molecular ground state wavefunctions based on a
superposition of non-orthogonal Slater determinants. The
symmetry-projected ansatz can account for static correlations in a
computationally efficient way. Here we present a variational extension
of this methodology applicable to excited states of the same symmetry
as the ground state.  Benchmark calculations on the C$_2$ dimer with a
modest basis set, which allows comparison with full configuration
interaction results, indicate that this extension provides a high
quality description of the low-lying spectrum for the entire
dissociation profile. We apply the same methodology to obtain the full
low-lying vertical excitation spectrum of formaldehyde, in good
agreement with available theoretical and experimental data, as well as
to a challenging model $C_{2v}$ insertion pathway for BeH$_2$. The
variational excited state methodology developed in this work has two
remarkable traits: it is fully black-box and will be applicable to
fairly large systems thanks to its mean-field computational cost.
\end{abstract}

\maketitle

\section{Introduction}
\label{sec:introduction}

The quantum mechanical character of a chemical system is reflected in
the discrete spectrum of electronic excitations. From the theoretical
point of view, the prediction of geometries and excitation energies
provides a means to interpret experimental electronic spectra. In
addition, optically forbidden states, which often play an important
role in the radiationless relaxation of a molecule, can be accessed.

When the excited state of interest has a different symmetry than the
ground state, one can use a ground-state formalism. In the
Hartree--Fock (HF) approximation, this approach is usually referred to
as the $\Delta$-SCF (self-consistent-field) method. On the other hand,
when the excited state of interest has the same symmetry as the
ground-state one has to resort to methods explicitly designed to treat
excited states.

Quantum chemical methods used to describe excited states can be
roughly categorized in two groups \cite{dreuw2005}. On the one hand,
high-quality wavefunction methods can be used to predict excitation
energies and oscillator strengths of low-energy transitions with great
accuracy. Among them we find methods based on general
multi-configuration SCF (MCSCF) and complete active-space SCF (CASSCF)
wavefunctions \cite{werner1981}, including the complete active-space
second-order perturbation theory (CASPT2)
\cite{roos1996}. Equation-of-motion and linear-response
coupled-cluster \cite{krylov2008,watts2008}, as well as
state-universal \cite{li2011} and state-specific \cite{ivanov2011}
multi-reference coupled cluster, approaches can also be used to
describe excited state properties. The symmetry-adapted-cluster
configuration interaction (SAC CI) \cite{nakatsuji1979} and Green's
function-based methods \cite{rohlfing2000} also deserve notice. All
these high-quality wavefunction approaches can be used in small
systems, although the meaning of ``small'' has been adapting to the
methodological and algorithmic advances seen in recent decades (see,
{\em e.g.}, Ref. \onlinecite{kus2009}). On the other hand, several
prominent methods can be used to access excited states at a reduced
computational cost, which permits the description of much larger
systems. The time-dependent density functional theory (TD-DFT)
\cite{casida2012} and configuration-interaction singles (CIS)
\cite{foresman1992} are perhaps the most widely used methods in this
category.

In this work, we describe yet another approach to describe excited
states of molecular systems by chains of variational calculations
based on symmetry-projected configurations. This approach, first
proposed by Schmid and co-workers \cite{schmid1986} has already proved
successful in the description of excited states in nuclear systems
\cite{schmid2004}. Recently, we have used the same approach to
describe ground and excited states of the two-dimensional periodic
Hubbard model \cite{rodriguez-guzman2012}. In the excited
symmetry-projected HF strategy, each state is described by (a set of)
symmetry-projected configurations. If the states are of the same
symmetry, the orthogonality between the states is enforced by a
modification of the ansatz. The advantages of this method are several:
\begin{itemize}
  \item The method can be regarded as having essentially mean-field
    computational cost.

  \item Unlike CIS or TD-DFT, the method can describe two-electron
    excitations with the same ease as one-electron processes.

  \item One does not need to compromise between the quality of the
    ground and excited states as is often done in state-averaged
    MCSCF approaches.

  \item Being a wavefunction ansatz, the evaluation of response
    properties and analytic derivative methods are in principle
    straightforward.
\end{itemize}

We show the potential of the method in providing a high-quality
low-lying spectrum of molecular systems. In particular, we focus on
the dissociation profile of the carbon dimer, which is challenging due
to the interaction between two low-lying states of the same
symmetry. Additionally, we discuss the application of the method to
compute the vertical excitation spectrum of formaldehyde. Due to its
simplicity, formaldehyde has been studied using a wide variety of
theoretical approaches, which facilitates the comparison with other
methods. Lastly, we consider a model of the insertion reaction of Be
into H$_2$, a challenging system commonly used to assess
state-of-the-art quantum chemical methods.

This work is organized as follows. In Sec. \ref{sec:formalism} we
describe in detail the formalism used in terms of symmetry-projected
HF configurations. In Sec. \ref{sec:comput_details} we briefly
describe our implementation of the method. We discuss in
Sec. \ref{sec:results} the application of the method to the
description of the dissociation profile of the carbon dimer, the
vertical excitation spectrum of formaldehyde, and the insertion
reaction of Be into H$_2$. Sec. \ref{sec:conclusions} is devoted to
concluding remarks.

\section{Formalism}
\label{sec:formalism}

We present in this section a detailed account of the formalism
employed in this work. We begin in Sec. \ref{ssec:sphf} by describing
the symmetry-projected HF ansatz for the ground state of a molecular
system with well defined quantum numbers. In Sec. \ref{ssec:excited}
we set out the excited symmetry-projected HF ansatz for states of the
same symmetry as the ground state, an approach introduced by Schmid
{\em et al}. \cite{schmid1986}. The variational optimization of the
considered wavefunctions is discussed in
Sec. \ref{ssec:optimization}. Lastly, in Sec. \ref{ssec:correlation},
we describe how one may go about building further correlations in both
the ground and excited states \cite{rodriguez-guzman2013}, even though
this is not something we have carried out in this work.

\subsection{Symmetry-projected Hartree--Fock}
\label{ssec:sphf}

In a seminal paper, L\"owdin \cite{lowdin1955b} introduced the
symmetry-projected HF ansatz for the ground state of a many-body
system of fermions. This is expressed as
\begin{equation}
  |\Psi \rangle = \hat{P} |\Phi \rangle,
\end{equation}
where $\hat{P}$ is a (set of) projection operator(s) that restores the
symmetries of a broken symmetry Slater determinant $|\Phi
\rangle$. This variational ansatz can account for strong correlations
due to spin or orbital degeneracies. It is important to stress that,
despite the multi-determinantal character in the wavefunction, the
ansatz above does not lose the connection to the single-particle
picture: the ansatz is fully determined by the set of molecular
orbitals occupied in $|\Phi \rangle$ \cite{lowdin1966}.

In the case of spin projection, L\"owdin suggested to use a projection
operator of the form
\begin{equation}
  \hat{P}^s = \prod_{l\neq s} \frac{\hat{S}^2 - l(l+1)}{s(s+1) -
    l(l+1)},
\end{equation}
where $s$ is used to label the quantum number to be recovered. The
projection operator is written as a product of two-body operators
rendering it impractical for routine calculations. Following work from
the nuclear physics community, we have discussed in
Ref. \onlinecite{jimenez-hoyos2012b} a more convenient form of the
projection operators used for spin and point-group symmetry
restoration. These are based on the forms introduced by Bayman
\cite{bayman1960} (number) and Villars \cite{villars1966} (angular
momentum). A similar and earlier spin-projection by rotation formalism
introduced by Percus and Rotenberg \cite{percus1962} has gone largely
unnoticed. For spin, we use projection-like operators of the form
\begin{equation}
  \hat{P}^s_{mk} = \frac{2s+1}{8\pi^2} \int d\Omega \, D^{s\ast}_{mk}
    (\Omega) \, \hat{R} (\Omega),
  \label{eq:spinproj}
\end{equation}
where $\Omega = (\alpha, \beta, \gamma)$ is the set of Euler angles
parametrizing the rotation in spin space, $D^s_{mk} (\Omega) \equiv
\langle s,m| \hat{R} (\Omega) |s,k \rangle$ is Wigner's $D$-matrix,
and $\hat{R} (\Omega)$ is the spin-rotation operator
\begin{equation}
  \hat{R} (\Omega) = \exp \left(-i \alpha \hat{S}_z \right) \,
                     \exp \left(-i \beta  \hat{S}_y \right) \,
                     \exp \left(-i \gamma \hat{S}_z \right).
\end{equation}
For more details about the form of the projection operators, we refer
the reader to Ref. \onlinecite{ring_schuck}. We note that in
Ref. \cite{jimenez-hoyos2012b} we incorrectly suggested that if $|\Phi
\rangle$ is a UHF-type Slater determinant, the projection operator is
simplified. While it is true that matrix elements (norm, Hamiltonian)
are simplified (ultimately, a single integration over $\beta$ is
required), the projection operator {\em does not} change.

In this work, we write the symmetry-projected HF ansatz in the form
\begin{equation}
  |\Psi_{j,m} \rangle = \sum_k f_k \, \hat{P}^j_{mk} |\Phi \rangle,
  \label{eq:sphf_ansatz}
\end{equation}
The subscripts $j,m$ in $|\Psi \rangle$ label the irreducible
representation and the row of the irrep to recover,
respectively \footnote{Here, $\hat{P}$ may stand for a product of
  projection operators for say, spin and point group restoration.}.
The form above is suitable for arbitrary non-Abelian symmetry groups,
including spin. The linear variational coefficients $\{ f \}$ are
introduced in order to remove unphysical dependencies of the energy
with respect to the orientation of the underlying state $|\Phi
\rangle$ \cite{schmid1984,ring_schuck}.

The energy associated with the symmetry-projected HF state of
Eq. \ref{eq:sphf_ansatz} is given by
\begin{align}
  E_j [\Phi]
  &= \, \frac{\sum_{kk'} f_k^\ast \, f_{k'} \, \langle \Phi|
    \hat{P}_{mk}^{j\dagger} \, \hat{H} \, \hat{P}^j_{mk'} |\Phi
    \rangle}{\sum_{kk'} f_k^\ast \, f_{k'} \, \langle \Phi|
    \hat{P}_{mk}^{j\dagger} \, \hat{P}^j_{mk'} |\Phi \rangle} \nonumber \\[4pt]
  &= \, \frac{\sum_{kk'} f_k^\ast \, f_{k'} \, \langle \Phi| \hat{H}
    \, \hat{P}^j_{kk'} |\Phi \rangle}{\sum_{kk'} f_k^\ast \, f_{k'} \,
    \langle \Phi| \hat{P}^j_{kk'} |\Phi \rangle},
  \label{eq:sphf_energy}
\end{align}
where we have used the properties of the projection operators
\cite{jimenez-hoyos2012b} and the fact that they commute with the
Hamiltonian. We have emphasized the independence of the energy
expression on the row of the irrep selected for non-Abelian groups.
We discuss in Appendix \ref{sec:matrix_elements} the evaluation of
norm and Hamiltonian overlaps between symmetry-projected
configurations.

In carrying out the optimization of the wavefunction ansatz of
Eq. \ref{eq:sphf_ansatz}, one can consider two possibilities:
\begin{itemize}
  \item In a projection-after-variation (PAV) approach, the
    broken-symmetry mean-field state $|\Phi \rangle$ is optimized
    variationally. The symmetry-projected energy is then computed in a
    single-shot evaluation.

  \item In a variation-after-projection (VAP) approach, the Slater
    determinant $|\Phi \rangle$ is optimized in the presence of the
    projection operators.
\end{itemize}
The PAV approach is appealing for its simplicity. However, it may lead
to unphysical behavior: dissociation profiles evaluated with the PAV
approach show derivative discontinuities at the point where the
broken-symmetry HF solution collapses back to the symmetry-adapted one
\cite{schlegel1986}.

The VAP approach is favored not only because it leads to lower
energies, but most importantly because the variation is performed for
the actual considered ansatz. As it will be shown below, optimizing
the state of Eq. \ref{eq:sphf_ansatz} in a VAP manner leads to
generalized Brillouin-like conditions that characterize the stationary
nature of the solution. A self-consistent VAP approach was the basis
of the extended Hartree--Fock method proposed by L\"owdin
\cite{lowdin1955b}. More often than not, EHF has been associated with
the use of a spin-projection operator on a reference unrestricted
determinant (the so-called spin-projected EHF \cite{mayer1980}).

Our previous work (Ref. \onlinecite{jimenez-hoyos2012b}) discussed the
self-consistent optimization of the symmetry-projected HF approach. In
this work, however, we follow a different strategy to carry out the
variational optimization, which we describe in more detail in
Sec. \ref{ssec:optimization}.

\subsection{The excited symmetry-projected HF approach}
\label{ssec:excited}

Having described the symmetry-projected HF approach for the
variational optimization of the ground state of a given symmetry, we
now turn our attention to excited states of the same symmetry as the
ground state. The excited symmetry-projected HF approach, which relies
on a Gram-Schmidt orthogonal construction, was introduced by Schmid
{\em et al}. \cite{schmid1986} in the nuclear physics community as the
{\em excited VAMP} (Variation After Mean-field Projection) strategy.

In order for a given ansatz to constitute a faithful representation of
an excited state, it must remain orthogonal to the ground state. In
variational strategies, this feature may be accomplished in two
alternative ways:
\begin{itemize}
  \item Use the same ansatz as the one employed in the ground state
    optimization. The orthogonality with respect to the ground state
    is enforced as a constraint; that is, one minimizes the Lagrangian
    \begin{equation}
      \mathcal{L} [\Psi] = E [\Psi] - \lambda \, \langle \Psi | \Psi_0
        \rangle,
    \end{equation}
    where $\lambda$ is a Lagrange multiplier and $|\Psi_0 \rangle$ is
    the ground state.

  \item Use an ansatz that is explicitly orthogonal to the ground
    state wavefunction. This is our preferred approach as the
    minimization problem remains unconstrained.
\end{itemize}

Let us assume that the symmetry-projected ground state is already
available. This we write as
\begin{align}
  |\Psi^0_{j,m} \rangle
  &\equiv \, |\psi^0_{j,m} \rangle
  \nonumber \\
  &= \, \sum_k f^0_k \, \hat{P}^j_{mk} |\Phi^0 \rangle,
\end{align}
where the $0$ superscript is used to denote the ground state
character. Our ansatz for the first excited state is given by
\begin{align}
  |\Psi^1_{j,m} \rangle
  &\equiv \, \left( 1 - \hat{S}^1 \right) |\psi^1_{j,m} \rangle
  \nonumber \\
  &= \, \left( 1 - \hat{S}^1 \right) \sum_k f^1_k \, \hat{P}^j_{mk}
    |\Phi^1 \rangle,
  \label{eq:1st_ansatz}
\end{align}
written in terms of the projector
\begin{equation}
  \hat{S}^1 = \frac{| \psi^0_{j,m} \rangle \, \langle \psi^0_{j,m}
    |}{\langle \psi^0_{j,m} | \psi^0_{j,m} \rangle},
\end{equation}
which guarantees the orthogonality of $|\Psi^1_{j,m} \rangle$ with
respect to the ground state. Here, the superscript $1$ is used to
denote that the first excited state is under consideration. The
variational flexibility in the ansatz of Eq. \ref{eq:1st_ansatz} lies
in the set of linear variational coefficients $\{ f^1 \}$ and the
Slater determinant $|\Phi^1 \rangle$, which is in general not
orthogonal to $|\Phi^0 \rangle$. Nonetheless, it is important to
stress that the actual wavefunction is not a single symmetry-projected
configuration but a linear combination.

A similar construction can be used for higher excited states. Having
the ground state and $q-1$ excited states already at our disposal, we
prepare an ansatz for the $q$-th excited state as
\begin{align}
  |\Psi^q_{j,m} \rangle
  &\equiv \, \left( 1 - \hat{S}^q \right) |\psi^q_{j,m} \rangle
  \nonumber \\
  &= \, \left( 1 - \hat{S}^q \right) \sum_k f^q_k \, \hat{P}^j_{mk}
    |\Phi^q \rangle,
  \label{eq:qth_ansatz}
\end{align}
with the projector $\hat{S}^q$ given by
\begin{align}
  \hat{S}^q &= \, \sum_{r,s=0}^{q-1} |\psi^r_{j,m} \rangle \, \left(
    A^{-1} \right)_{rs} \, \langle \psi^s_{j,m}|,
  \label{eq:qth_proj} \\
  A_{rs} &= \, \langle \psi^r_{j,m} | \psi^s_{j,m} \rangle.
  \label{eq:defAmat}
\end{align}
The projector $\hat{S}^q$ guarantees orthogonality with respect to the
ground state and the $q-1$ excited states previously considered. We
note that, along with the linear coefficients $\{ f^q \}$, a single
Slater determinant $|\Phi^q \rangle$ determines the full flexibility
in the ansatz of Eq. \ref{eq:qth_ansatz}. The energy functional
associated with the $q$-th excited state wavefunction becomes
\begin{equation}
  E^q_j [\{ f^q \}, |\Phi^q \rangle] =
  \frac{\sum_{kk'} f_k^{q\ast} \, f_{k'}^q \, \mathcal{H}^q_{kk'}}
  {\sum_{kk'} f_k^{q\ast} \, f_{k'}^q \, \mathcal{N}^q_{kk'}},
  \label{eq:qth_energy}
\end{equation}
where $E^q_j$ is the energy of the $q$-th excited state. Here, the
matrices $\mathcal{N}^q$ and $\mathcal{H}^q$ are given by
\begin{subequations}
  \begin{align}
    \mathcal{N}^q_{kk'} &= \,
    \langle \Phi^q | \hat{P}^j_{km} \, \left( 1 - \hat{S}^q \right) \,
      \hat{P}^j_{mk'} | \Phi^q \rangle,
    \label{eq:qth_overlap} \\
    \mathcal{H}^q_{kk'} &= \,
    \langle \Phi^q | \hat{P}^j_{km} \, \left( 1 - \hat{S}^q \right) \,
      \hat{H} \, \left( 1 - \hat{S}^q \right) \, \hat{P}^j_{mk'} |
      \Phi^q \rangle.
    \label{eq:qth_hamilt}
  \end{align}
\end{subequations}
We note that even if it may appear otherwise, all matrix elements in
the energy functional of Eq. \ref{eq:qth_energy} can be evaluated in
terms of norm and Hamiltonian overlaps between symmetry-projected
configurations (using a single projection operator). For instance,
\begin{align}
  \mathcal{N}^q_{kk'}
  &= \, \langle \Phi^q | \hat{P}^j_{km} \, \hat{P}^j_{mk'} | \Phi^q
    \rangle - \sum_{r,s=0}^{q-1} \sum_{ll'} \langle \Phi^q |
    \hat{P}^j_{km} \, \hat{P}^j_{ml} | \Phi^r \rangle \, \left( A^{-1}
    \right)_{rs} \, \langle \Phi^s | \hat{P}^j_{l'm} \, \hat{P}^j_{mk'}
    | \Phi^q \rangle \, f_l^r \, f_{l'}^{s\ast}
  \nonumber \\
  &= \, \langle \Phi^q | \hat{P}^j_{kk'} | \Phi^q \rangle -
    \sum_{r,s=0}^{q-1} \sum_{ll'} \langle \Phi^q | \hat{P}^j_{kl} |
    \Phi^r \rangle \, \left( A^{-1} \right)_{rs} \, \langle \Phi^s |
    \hat{P}^j_{l'k'} | \Phi^q \rangle \, f_l^r \, f_{l'}^{s\ast}.
  \nonumber
\end{align}
It follows that all states in the irreducible representation $j$
thereby obtained are degenerate. The expressions used to evaluate
matrix elements between symmetry-projected configurations are provided
in Appendix \ref{sec:matrix_elements}. The variational optimization of
the energy functional of Eq. \ref{eq:qth_energy} is considered in the
following subsection.

Let us assume that, through the scheme described above, the ground
state and all $q$ excited states have already been obtained. These
states, $\{ |\Psi^r_{j,m} \rangle \, | \, r = 0, \ldots, q \}$, are
orthogonal among themselves, but they are not necessarily orthogonal
through the Hamiltonian. One can therefore carry out a diagonalization
of the Hamiltonian in this basis or, equivalently, in the basis of $\{
|\psi^r_{j,m} \rangle \}$. The eigenvalue equations can be written as
\begin{equation}
  B \, g = g \, A \, \varepsilon,
  \label{eq:final_diag}
\end{equation}
where $g$ is the matrix of eigenvectors, $\varepsilon$ is the diagonal
matrix of eigenvalues, $A$ is defined by Eq. \ref{eq:defAmat}, and
\begin{equation}
  B_{rs} = \langle \psi^r_{j,m} | \hat{H} | \psi^s_{j,m} \rangle.
\end{equation}
In this way, the states $\{ |\psi^r_{j,m} \rangle \}$ are allowed to
interact through the Hamiltonian. The states obtained $\{
|\eta^r_{j,m} \rangle \, | \, r = 0, \ldots, q \}$ through the
diagonalization, expressed as
\begin{align}
  |\eta^r_{j,m} \rangle
  &= \, \sum_{s=0}^q g_{sr} |\psi^s_{j,m} \rangle
  \nonumber \\
  &= \, \sum_{s=0}^q g_{sr} \sum_k f^s_k \, \hat{P}^j_{mk} |\Phi^s
    \rangle,
\end{align}
are orthogonal through the Hamiltonian and thus represent a faithful
representation of the low-lying spectrum of the considered
symmetry. Note also that the final diagonalization of
Eq. \ref{eq:final_diag} can account for further correlations in the
ground state, that is, beyond those described by the
symmetry-projected HF ansatz.

\subsection{Variational optimization}
\label{ssec:optimization}

We proceed to discuss the strategy we use to variationally optimize
wavefunctions based on symmetry-projected configurations. Without loss
of generality, we work on the optimization of the $q$-th excited state
wavefunction, whose associated energy functional is given by
Eq. \ref{eq:qth_energy}. The ground state optimization can be carried
out in a similar fashion. The optimization has to be performed with
respect to the set of linear variational coefficients $\{ f^q \}$ and
with respect to the underlying determinant $|\Phi^q \rangle$.

The variation of the energy functional (Eq. \ref{eq:qth_energy}) with
respect to $\{ f^{q\ast} \}$ leads to the generalized eigenvalue
problem
\begin{equation}
  \sum_{k'} \left( \mathcal{H}^q_{kk'} - E^q_j \, \mathcal{N}^q_{kk'}
    \right) f^q_{k'} = 0 \qquad \forall \qquad k,
\end{equation}
which has to be solved subject to the normalization constraint
\begin{equation}
  f^{q\dagger} \, \mathcal{N}^q \, f^q = \mathbf{1}_{d'},
\end{equation}
where $d' \leq d$ and $d$ is the dimension of the irreducible
representation recovered by the projection. The Hamiltonian
$\mathcal{H}^q$ and overlap $\mathcal{N}^q$ matrices are given by
Eqs. \ref{eq:qth_hamilt} and \ref{eq:qth_overlap}, respectively. In
addition, $E^q_j$ is the lowest-energy solution to the eigenvalue
problem (it constitutes the energy of the $q$-th excited state); all
other solutions are discarded at this point.

The variation of the energy functional with respect to the underlying
determinant, $|\Phi^q \rangle$, is more convoluted. We use a
parametrization based on the Thouless theorem, which states that the
$N$-electron Slater determinant $|\Phi^q \rangle$ can be written in
terms of another (reference) $N$-electron Slater determinant $|\Phi
\rangle$ as
\begin{align}
  |\Phi^q \rangle &= \, \eta \, \exp (\hat{Z}^q) |\Phi
    \rangle,
  \label{eq:thou_q} \\
  \hat{Z}^q &= \, \sum_{ph} Z^q_{ph} \, b_p^\dagger \, b_h,
  \label{eq:thou_zph}
\end{align}
as long as $|\Phi^q \rangle$ is not orthogonal to $|\Phi
\rangle$. Here, $\eta = \langle \Phi^q | \Phi \rangle$ is a
normalization factor, and the sum in Eq. \ref{eq:thou_zph} is over
particle and hole operators defined by the orbitals characterizing
$|\Phi \rangle$. The coefficients $Z^q_{ph}$ are unique.

The Thouless theorem permits an efficient parametrization of the
Slater determinant $|\Phi^q \rangle$. That is, we use
Eq. \ref{eq:thou_q} and treat the coefficients $Z^q_{ph}$ as
variational parameters. We note that this Thouless parametrization has
not been frequently used in chemistry. Mang \cite{mang1975}, among
others, suggested its use in the nuclear physics community in the
context of a Hartree--Fock--Bogoliubov reference vacuum. Recently,
Noga and {\v S}imunek \cite{noga2010} used a Thouless matrix, in a
unitary coupled-cluster singles framework, to carry out the
optimization of independent particle model wavefunctions. Their
approach is similar in spirit to ours, though the actual algorithm has
important differences. A Thouless-based optimization is also closely
related to the quadratically-convergent algorithm suggested by Backsay
\cite{backsay1981}. We point the interested reader to
Refs. \onlinecite{egido1995,jimenez-thesis} for a more detailed
description of the approach we use.

Using Eq. \ref{eq:thou_q}, we can write the energy functional of
Eq. \ref{eq:qth_energy} as one depending on the coefficients
$Z^q_{ph}$,
\begin{equation}
  E^q_j [\{ f^q \}, Z^q] = \frac{\displaystyle \sum_{kk'} f_k^{\ast q} \,
    f_{k'}^q \, \langle \Phi | \exp (\hat{Z}^{q\dagger}) \,
    \hat{P}^j_{km} \, \left( 1 - \hat{S}^q \right) \, \hat{H} \,
    \left( 1 - \hat{S}^q \right) \, \hat{P}^j_{mk'} \exp (\hat{Z}^q) |
    \Phi \rangle}{\displaystyle \sum_{kk'} f_k^{\ast q} \, f_{k'}^q \,
    \langle \Phi | \exp (\hat{Z}^{q\dagger}) \, \hat{P}^j_{km} \,
    \left( 1 - \hat{S}^q \right) \, \hat{P}^j_{mk'} \exp (\hat{Z}^q) |
    \Phi \rangle},
  \label{eq:qth_energy-par}
\end{equation}
where $|\Phi \rangle$ is an arbitrary reference state used for the
minimization.

A stationary point of the energy functional of
Eq. \ref{eq:qth_energy-par} is reached when the energy gradient, given
by
\begin{align}
  G^q_{ph}
  &\equiv \, \frac{\partial}{\partial Z^{q\ast}_{ph}} \, E^q_j [\{ f^q
    \}, Z^q] \nonumber
  \\
  &= \, \frac{\displaystyle \sum_{kk'} f_k^{\ast q} \, f_{k'}^q \,
    \langle \Phi^q | b_h^\dagger \, b_p \, \hat{P}^j_{km} \, \left( 1
    - \hat{S}^q \right) \, \left( \hat{H} - E^q_j \right) \left( 1 -
    \hat{S}^q \right) \, \hat{P}^j_{mk'} |\Phi^q
    \rangle}{\displaystyle \sum_{kk'} f_k^{\ast q} \, f_{k'}^q \,
    \langle \Phi^q | \hat{P}^j_{km} \, \left( 1 - \hat{S}^q \right) \,
    \hat{P}^j_{mk'} |\Phi^q \rangle},
\end{align}
vanishes for all elements of $G^q$. Here, the HF operators
$b_h^\dagger$ and $b_p$ are associated with the determinant $|\Phi
\rangle$ (and not $|\Phi^q \rangle$). This is sometimes referred to as
the global gradient. The local gradient $\mathcal{G}^q$ at $|\Phi
\rangle = |\Phi^q \rangle$, given by
\begin{align}
  \mathcal{G}^q_{ph}
  &\equiv \, \left. \frac{\partial}{\partial Z^{q\ast}_{ph}} \, E^q_j
    [\{ f^q \}, Z^q] \right|_{Z^q_{ph} = 0} \nonumber
  \\
  &= \, \frac{\displaystyle \sum_{kk'} f_k^{\ast q} \, f_{k'}^q \,
    \langle \Phi^q | b_h^{q\dagger} \, b_p^q \, \hat{P}^j_{km} \,
    \left( 1 - \hat{S}^q \right) \, \left( \hat{H} - E^q_j \right)
    \left( 1 - \hat{S}^q \right) \, \hat{P}^j_{mk'} |\Phi^q
    \rangle}{\displaystyle \sum_{kk'} f_k^{\ast q} \, f_{k'}^q \,
    \langle \Phi^q | \hat{P}^j_{km} \, \left( 1 - \hat{S}^q \right) \,
    \hat{P}^j_{mk'} |\Phi^q \rangle},
\end{align}
in which HF operators associated with $|\Phi^q \rangle$ are used, can
be related to the global gradient by \cite{jimenez-hoyos2012}
\begin{equation}
  G^q = \tilde{L}^{\trans -1} \, \mathcal{G}^q \, L^{\ast -1}.
\end{equation}
Here, $\tilde{L}$ and $L$ are $(M-N) \times (M-N)$ and $N \times N$
matrices, respectively, obtained from standard Cholesky decompositions
(see Ref. \onlinecite{jimenez-hoyos2012}). We note that the local
gradient also vanishes at a stationary point of the energy functional.

Once the optimal $|\Phi^q \rangle$ has been found, it is convenient to
have a unique representation of the molecular orbitals characterizing
the Slater determinant (recall that the functional is invariant to
unitary transformations among the occupied orbitals). This can be
accomplished by diagonalizing the $H^{11}$ sector of the Hamiltonian,
whose hole-hole and particle-particle blocks are given by
\begin{subequations}
  \begin{align}
    H^{11}_{hh'} = \langle \Phi^q | b_h^\dagger \, H \, b_{h'}^\dagger
      | \Phi^q \rangle - \delta_{hh'} \langle \Phi^q | \hat{H} | \Phi^q
      \rangle, \\
    H^{11}_{pp'} = \langle \Phi^q | b_p \, H \, b_{p'}^\dagger |
      \Phi^q \rangle - \delta_{pp'} \langle \Phi^q | \hat{H} | \Phi^q
      \rangle.
  \end{align}
\end{subequations}
Note that this is simply a way of finding semi-canonical orbitals,
using traditional quantum chemical jargon.

We close this section by listing some of the advantageous features
that a variational optimization based on a Thouless parametrization
provides:
\begin{itemize}
  \item The minimization problem is unconstrained, with as many
    parameters as linearly independent variables. Powerful algorithms
    (conjugate gradient, quasi-Newton methods) for unconstrained
    minimization can be used \cite{nocedal_wright}.

  \item Because of the gradient-based approach used, one is guaranteed
    that the optimization will either converge to a stationary point
    within a specified tolerance or the algorithm used will fail.

  \item The application of the method to symmetry-projected approaches
    or arbitrary wavefunctions expressed in terms of Slater
    determinants is straightforward.

  \item The method does not require one to {\em a priori} decide how
    to occupy the orbitals, which a diagonalization approach
    requires. For HF, an {\em aufbau} occupation leads to the lowest
    energy solution, but the same need not be true for more general
    functionals.
\end{itemize}

\subsection{Correlations in the ground and excited states}
\label{ssec:correlation}

In the previous sections, we have considered an ansatz for the ground
and excited states of a given symmetry. Each state is described by
essentially a single symmetry-projected HF configuration. If this
description proves insufficient, one can consider a more general
ansatz written as a linear combination of symmetry-projected
configurations as a trial wavefunction for each state. This approach
has been used to describe ground-state correlations of molecular
systems in Ref. \onlinecite{jimenez-hoyos2013} and in the Hubbard
model in Ref. \onlinecite{rodriguez-guzman2013}. We briefly describe
the idea in this section, even though we do not include results from
such multi-component approach in our calculations.

In a multi-component approach, the ground state is expanded as a
linear combination of symmetry-projected configurations
\begin{align}
  |\Psi^0_{j,m} \rangle
  &\equiv \, |\psi^0_{j,m} \rangle
  \nonumber \\
  &= \, \sum_k \hat{P}^j_{mk} \sum_{z=1}^{n_0} f^0_{z;k} |\Phi^0_z
    \rangle.
\end{align}
Here, once again the superscript $0$ denotes the ground state
character of the considered ansatz. The trial wavefunction is expanded
as a linear combination of $n_0$ symmetry-projected configurations,
obtained from the corresponding set of (non-orthogonal) Slater
determinants $\{ |\Phi^0_z \rangle \, | \, z = 1, \ldots, n_0 \}$.

The ansatz for the $q$-th excited state is similar to that from the
single-configuration approach. It is given by
\begin{align}
  |\Psi^q_{j,m} \rangle
  &\equiv \, \left( 1 - \hat{S}^q \right) |\psi^q_{j,m} \rangle
  \nonumber \\
  &= \, \left( 1 - \hat{S}^q \right) \sum_k \hat{P}^j_{mk}
    \sum_{z=1}^{n_q} f^q_{z;k} |\Phi^q_z \rangle,
  \label{eq:qth_mr_ansatz}
\end{align}
with the projector $\hat{S}^q$ given by an expression analogous to
Eq. \ref{eq:qth_proj}. (Note, nonetheless, that each state
$|\psi^r_{j,m} \rangle$ is given by a linear combination of $n_r$
symmetry-projected configurations.)

One may now wonder how the variational optimization is performed in
this multi-component approach. The two extreme strategies are:
\begin{itemize}
  \item All the determinants $\{ |\Phi^q_z \rangle \, | \, z = 1,
    \ldots, n_q \}$ describing the $q$-th excited state are optimized
    at once. This is known in the literature as the resonating
    Hartree--Fock approach (Res HF), first introduced by Fukutome
    \cite{fukutome1988}.

  \item A step-wise construction is used in which only the last added
    determinant is optimized while the previously obtained remain
    frozen. This is known, in the nuclear physics community, as the
    few-determinant (FED) approach introduced by Schmid {\em et
      al}. \cite{schmid1989}. In combination with the Gram-Schmidt
    orthogonal construction used for the excited states, it is
    referred to as the {\em excited FED VAMP} strategy
    \cite{schmid1989}. We note that a similar approach, even if Slater
    determinants were used in place of the symmetry-projected
    configurations, was employed by Koch and Dalgaard \cite{koch1993}
    in ground state optimizations.
\end{itemize}

We refer the reader to our recent work on the one-dimensional Hubbard
Hamiltonian \cite{rodriguez-guzman2013} where the merits of this
approach have been discussed.

\section{Computational details}
\label{sec:comput_details}

We have developed a computer program that can optimize
symmetry-projected HF states (as well as excited states) using a
Thouless parametrization, as described in
Sec. \ref{ssec:optimization}. This is different from our original work
(see Ref. \onlinecite{jimenez-hoyos2012b}), which used a
diagonalization based approach. A limited-memory
Broyden--Fletcher--Goldfarb--Shanno (BFGS) \cite{nocedal1980,liu1989}
quasi-Newton method is used as the unconstrained minimization
algorithm. The program interfaces with the \verb+GAUSSIAN+ suite
\cite{gaussian} to retrieve one- and two-electron integrals. Our
program is parallelized (MPI-based) over the grid to perform the
symmetry restoration (spatial and/or spin). We note that if a single
symmetry-projected configuration is used to describe each state (as
done in this work), the excited method scales linearly with the order
of the state described. That is, the optimization of the first excited
state is twice as expensive as that of the ground
state. Unfortunately, we have not yet implemented the capability of
evaluating oscillator strengths of the excited states, but it is
straightforward to do so once the appropriate integrals are
available. We prepare an initial guess for the broken symmetry
determinants by taking the converged HF solution and mixing a few
orbitals closest to the Fermi energy using a randomly prepared unitary
matrix. A similar strategy has been used in our recent study of the
one-dimensional Hubbard model \cite{rodriguez-guzman2013}.

\section{Results and discussion}
\label{sec:results}

We discuss the application of the excited symmetry-projected HF method
to three different systems: the dissociation profile of the carbon
dimer, the vertical excitation spectrum of formaldehyde, and a model
for the insertion reaction of Be in H$_2$. We consider simple systems
in small basis in order to compare with previously reported
results. Lastly, we emphasize that, in order to showcase the excited
symmetry-projected HF strategy, it has to be applied to systems for
which a few excited states of the same symmetry are of interest.

\subsection{Dissociation profile of the carbon dimer}
\label{ssec:carbon_dimer}

We consider the dissociation profile of the carbon dimer in the
6-31G(d) basis, for which the full CI (FCI) dissociation profile was
reported by Abrams and Sherrill in Ref. \cite{abrams2004}. The correct
description of the dissociation profile of the carbon dimer is quite
challenging from a theoretical point of view: not only is a
double-bond being broken, but there is a low-lying excited state of
the same-symmetry ($^1 \Sigma_g^+$) as the ground state nearby in
energy. In fact, an avoided crossing occurs at $\approx 1.7 \,
\mathring{\mathrm{A}}$, where the character of the two states is
interchanged. In addition, there is also a low-lying $^1 \Delta_g$
state that becomes the ground state at large interatomic separation. A
further complication arises because, as described by Abrams and
Sherrill, within the $D_{2h}$ subgroup available in most quantum
chemical packages, the $^1 \Sigma_g^+$ and the $^1 \Delta_g$ states
have the same $^1 A_g$ symmetry \cite{abrams2004}.

An assessment of the ability of several sophisticated quantum chemical
methods to describe the dissociation profile was presented in
Refs. \cite{abrams2004} and \cite{sherrill2005}. Most coupled-cluster
approaches fail to provide even a qualitatively correct description of
the dissociation profile of the ground state, with its characteristic
non-Morse-like behavior due to the avoided crossing. Only
multi-reference approaches such as CASPT2 or multi-reference CI
\cite{sherrill2005} can accurately describe the dissociation profile of
all three states considered.

\begin{figure}[!htbp]
  \centering
  \includegraphics[width=0.85\textwidth]{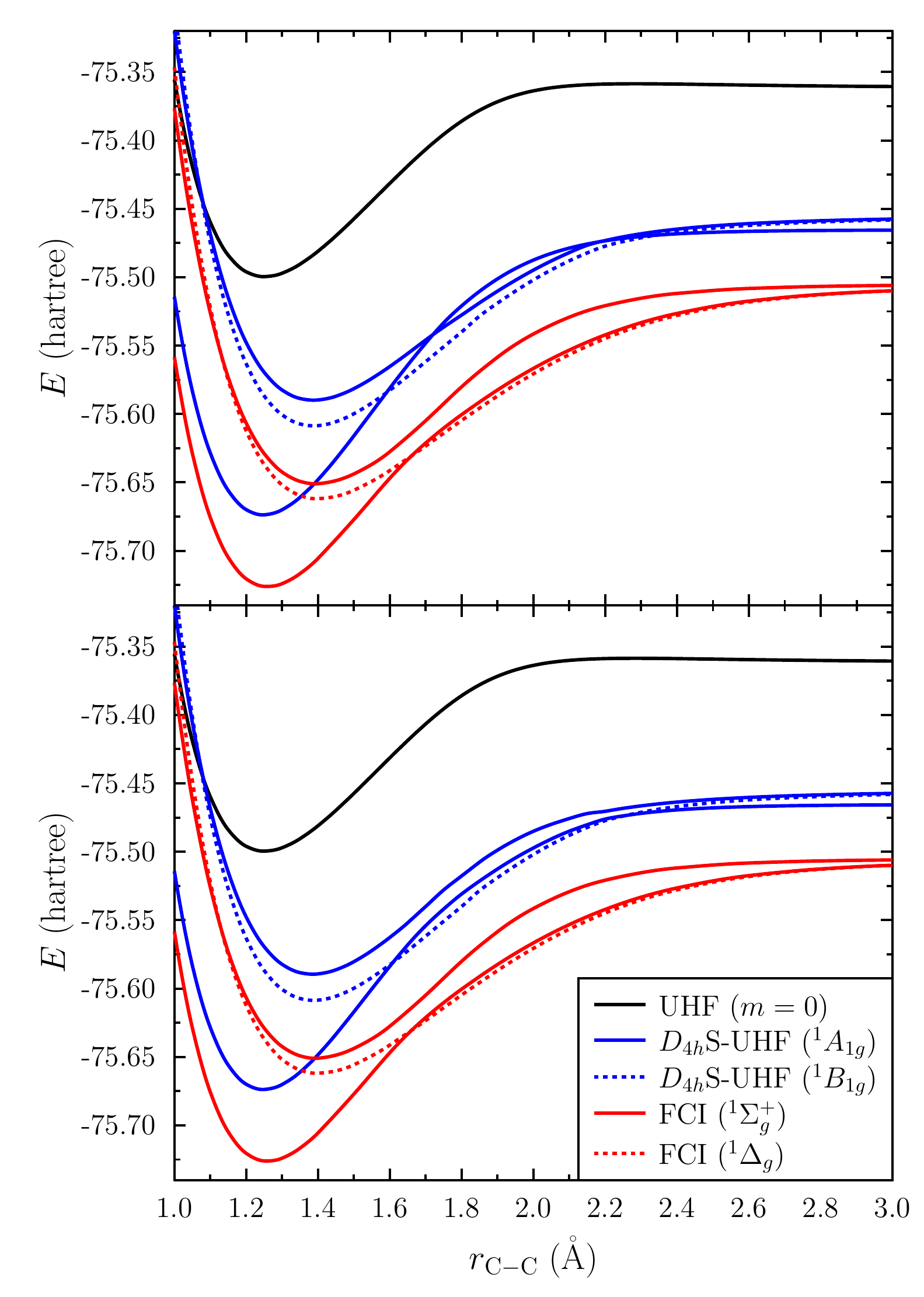}
  \caption{Dissociation profiles for low-lying singlet states of the
    C$_2$ molecule computed with the $D_{4h}$S-UHF / 6-31G(d)
    method. A comparison with FCI curves from Ref. \cite{abrams2004}
    is shown. The $D_{4h}$S-UHF profiles for the two $^1 A_{1g}$
    states as obtained before (top panel) and after (bottom panel) the
    final diagonalization (Eq. \ref{eq:final_diag}) are displayed. The
    avoided crossing is correctly described after the two $^1 A_{1g}$
    states are allowed to interact through the Hamiltonian.}
  \label{fig:c2dissoc}
\end{figure}

The dissociation profile of four low-lying singlet states of C$_2$ as
predicted with the excited $D_{4h}$S-UHF method is shown in
Fig. \ref{fig:c2dissoc}. Here, the $D_{4h}$S-UHF notation implies that
spin projection (S) and spatial symmetry-restoration (in the $D_{4h}$
point group) have been broken and restored from an underlying
determinant of unrestricted HF (UHF) character. The use of the
$D_{4h}$ subgroup allows us to distinguish between the $^1 \Sigma_g^+$
and the $^1 \Delta_g$ irreducible representations of the $D_{\infty
  h}$ group: the $^1 \Delta_g$ state transforms as the $^1 B_{1g}$
irreducible representation in the $D_{4h}$ subgroup. For the two $^1
A_{1g}$ states we show the profiles obtained before (top panel) and
after (bottom panel) they are allowed to interact through the
Hamiltonian in the final diagonalization of Eq. \ref{eq:final_diag}.

Several features of the exact dissociation profile are correctly
described with our $D_{4h}$S-UHF approximation. In particular, we
observe that the $B_{1g}$ state is correctly predicted to be the
lowest energy state for $r_{\mathrm{C-C}} > 1.6 \,
\mathring{\mathrm{A}}$. The avoided crossing observed in the FCI
profile appears after the two $^1 A_{1g}$ symmetry-projected
configurations are allowed to interact. In this way, the dissociation
profile predicted for the lowest-lying $^1 A_{1g}$ state correctly
displays the characteristic non-Morse-like behavior. One should note,
however, that the carbon-carbon distance of closest approach between
the two $^1 A_{1g}$ states is slightly larger than in the FCI
solution. We finally point out that the curves obtained for the three
states for which the FCI solution is available are fairly parallel to
the latter. This validates our description of the ground and excited
states of C$_2$ in terms of symmetry-projected configurations. We
emphasize that essentially a single symmetry-projected configuration
was used for each state: two symmetry-projected configurations were
used to describe two states of $^1 A_{1g}$ symmetry.

\subsection{Vertical excitation spectrum of formaldehyde}

Formaldehyde is the simplest of the carbonyl compounds and as such it
is ubiquitous in nature. The presence of a $\pi$-electron system and
the lone pairs of oxygen permit $n \rightarrow \pi^\ast$ and $\pi
\rightarrow \pi^\ast$ valence transitions, making formaldehyde
photochemically active.  Because of its small size and availability,
formaldehyde has been widely studied both experimentally and
theoretically.

Some vertical excitation transitions of formaldehyde have been
experimentally determined (see Refs \cite{robin} and
\cite{taylor1982}). The vertical excitation spectrum of formaldehyde
has been studied theoretically by several authors, both to help in the
assignment of the spectrum, as well as to test different theoretical
approaches. We refer the reader to the work by Hadad {\em et
  al}. \cite{hadad1993}, Pitarch-Ruiz {\em et al}. \cite{pitarch2003},
and references therein. Recently, Schreiber {\em et
  al}. \cite{schreiber2008} have provided best theoretical estimates
for some low-energy valence and Rydberg transitions of formaldehyde.

Our focus here is to test whether our approach can provide reasonable
excitation energies. In particular, we focus on the vertical
excitation spectrum as we currently lack the ability to optimize the
geometries of ground and excited states. We use the ground state
$C_{2v}$ geometry from Ref. \cite{hadad1993} (optimized with
MP2/6-31G(d)). The basis set 6-311(2+,2+)G(d,p) we use was also
obtained from the same work. The second set of diffuse functions was
found necessary in order to correctly describe the Rydberg transitions
at the CIS level.

\begin{figure}[!htbp]
  \centering
  \includegraphics[width=0.85\textwidth]{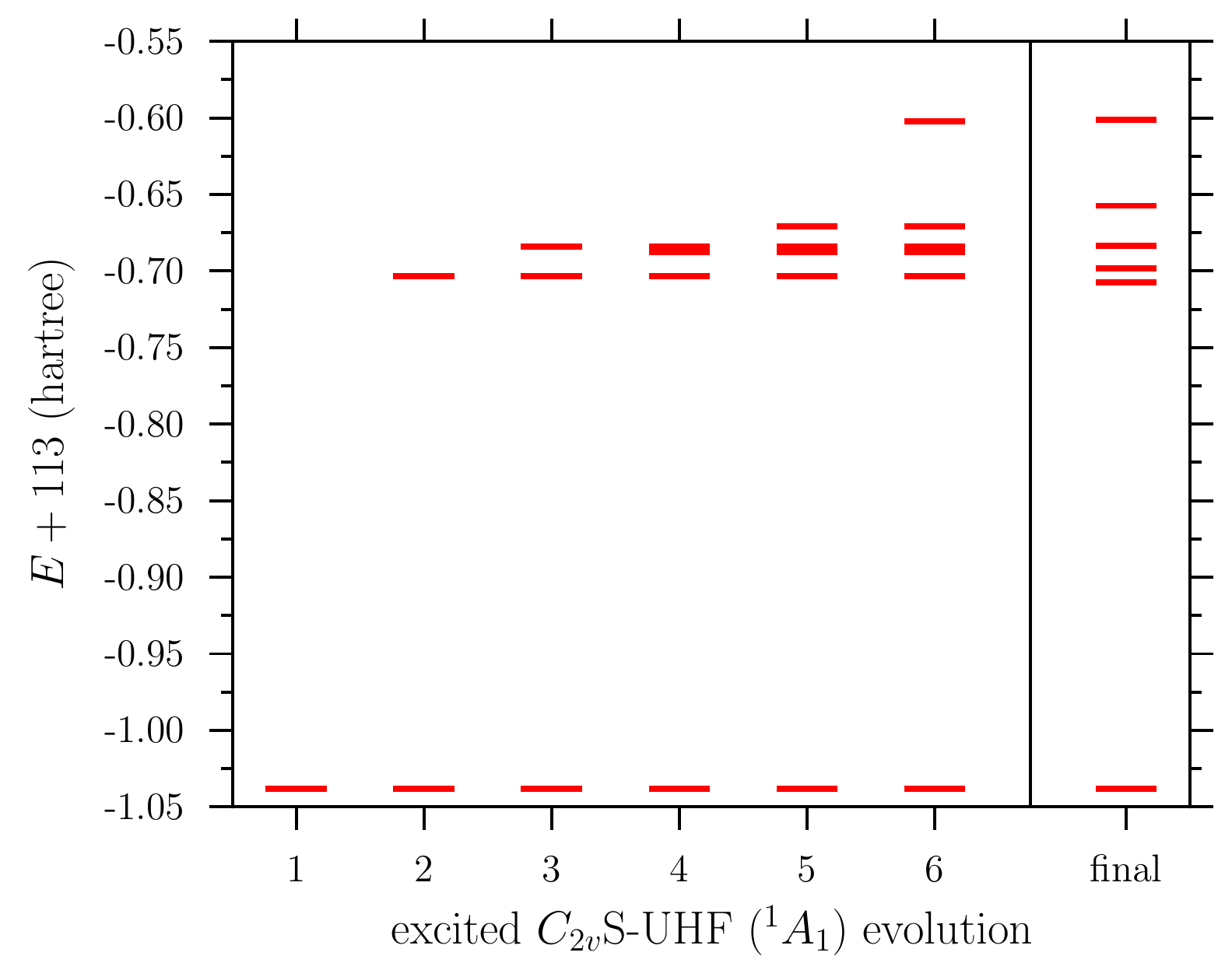}
  \caption{Evolution of the $^1 A_1$ spectrum of formaldehyde as
    computed with the excited $C_{2v}$S-UHF method with increasing
    number of symmetry-projected configurations. The last column shows
    the spectrum obtained after the final diagonalization of
    Eq. \ref{eq:final_diag} is carried out.}
  \label{fig:formal-evol}
\end{figure}

In Fig. \ref{fig:formal-evol} we show how six different singlet $A_1$
states are obtained by the chain of variational calculations defined
in the excited symmetry-projected ($C_{2v}$S-UHF) approach. Here, the
notation $C_{2v}$S-UHF implies that spin and spatial symmetry (in the
$C_{2v}$ framework) is restored from a broken symmetry UHF-type
determinant. Observe that the states are not necessarily obtained in a
strict increasing-energy order. In our $C_{2v}$S-UHF calculations for
triplet states, we have used an $m_s=1$ UHF-type determinant; the use
of $m_s=0$ determinants would lead to different
results \footnote{This, however, does not imply that all the triplet
  components in a symmetry-projected construction are not
  degenerate. If a UHF-type determinant $|\Phi \rangle$ has $m_s = z$,
  then all $2j+1$ states of the form $\hat{P}^j_{mz} \, |\Phi
  \rangle$, with $m=-j, \ldots, +j$, are degenerate.}. The right-most
column shows the resulting set of states after the final
diagonalization of Eq. \ref{eq:final_diag}. In this particular case,
the ground state gains almost no additional correlations as it is well
separated from other states energetically. On the other hand, several
of the states interact strongly as evidenced by the large differences
observed from column 6 to the column labeled as ``final''.

\begin{figure}[!htbp]
  \centering
  \includegraphics[width=0.85\textwidth]{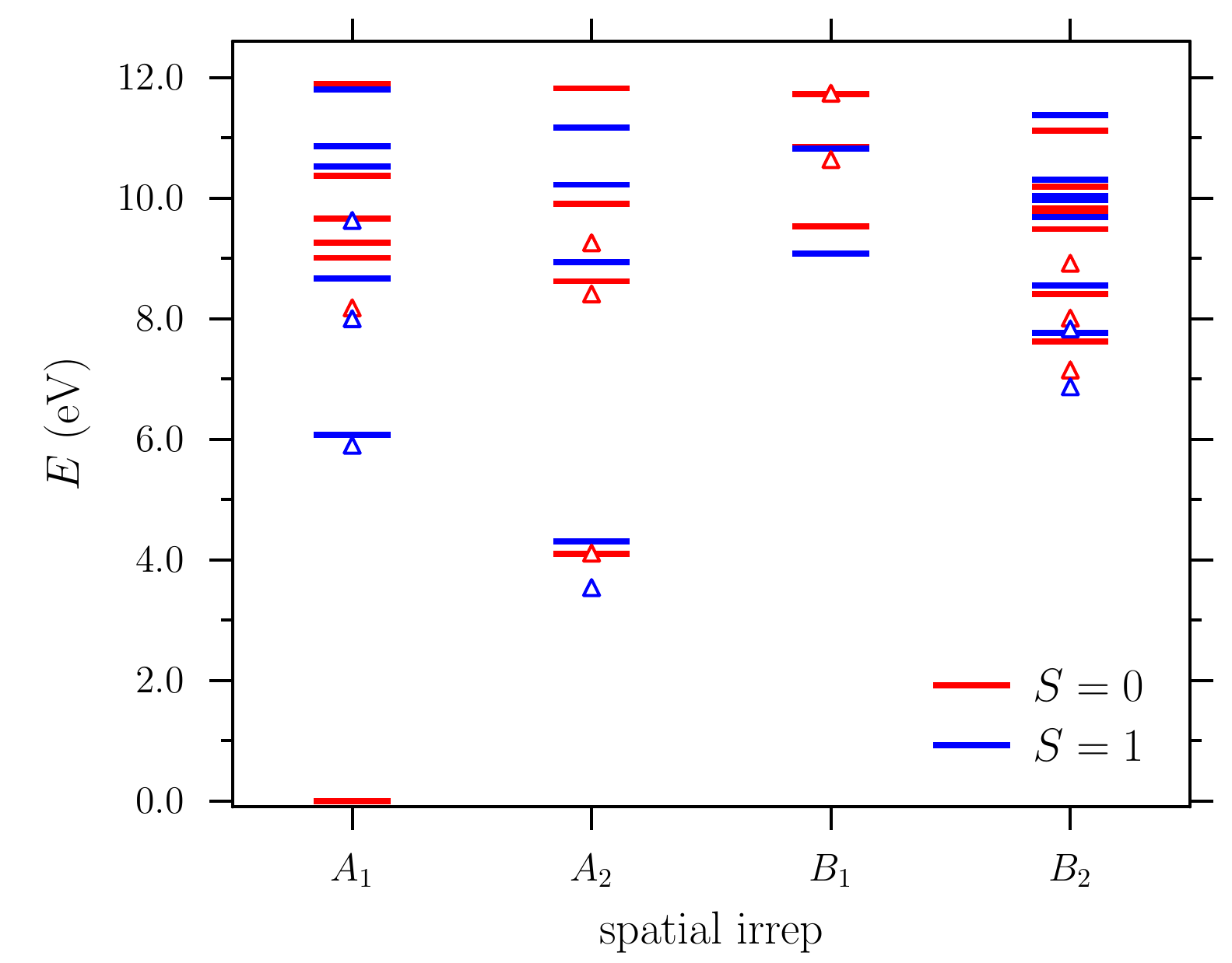}
  \caption{Low-lying singlet and triplet states of the formaldehyde
    molecule predicted with the $C_{2v}$S-UHF method. The
    6-311(2+,2+)G(d,p) basis was used in the
    calculations. Experimental excitation energies from
    Refs. \onlinecite{robin,taylor1982,hadad1993} are shown as red and
    blue triangles for singlet and triplet states, respectively.}
  \label{fig:formal-spectrum}
\end{figure}

We show, in Fig. \ref{fig:formal-spectrum} the full low-lying singlet
and triplet vertical spectrum of formaldehyde predicted with the
$C_{2v}$S-UHF approach. A comparison with experimental results from
Refs. \onlinecite{robin,taylor1982} and a few other results compiled
in Ref. \onlinecite{hadad1993} is also provided. As we have used a
limited basis set and our treatment of electron correlation is only
approximate, we cannot expect perfect agreement with experimental
numbers. The agreement between our $C_{2v}$S-UHF and the experimental
excitation energies (for both singlet and triplet states) is
remarkable, as each of the states obtained is described by essentially
a single symmetry-projected configuration. There is no {\em a priori}
reason to expect that all states should be well approximated by a
single symmetry-projected configuration, or that the quality obtained
for the different irreducible representations should be
comparable. Nevertheless, the agreement with the experimental
excitation energies is quite good, with maximum deviations of $\approx
1$ eV for both singlet and triplet states.

We show in Table \ref{tab:formaldehyde} a comparison of the predicted
low-lying vertical excitation of formaldehyde with $C_{2v}$S-UHF and
other results available from the literature. In particular, we compare
with the CIS and CIS-MP2 results of Ref. \cite{hadad1993}, where the
latter includes an electron correlation correction to the CIS energies
via perturbation theory through second order, and with the (SC)$^2$
multi-reference (MR) CI with singles and doubles (SD) of
Ref. \cite{pitarch2003}, which constitute best available theoretical
estimates. The (SC)$^2$ scheme is a self-consistent dressing procedure
that, among other effects, corrects the size-extensivity of MR-CI
\cite{daudey1993}. The CIS and CIS-MP2 calculations use the same basis
set and geometry that we have used. The MR-CISD calculations use a
large atomic natural-orbital-type [$6s5p3d2f/4s3p2d$] basis for C,O/H
augmented with a $3s3p3d$ adapted Rydberg series. The latter diffuse
functions were placed in the charge center of the $^2 B_2$ state of
the formaldehyde cation. The ground-state geometry used in MR-CISD
calculations was described in Ref. \onlinecite{pitarch2003}; it
deviates $\approx 0.01 \mathring{\mathrm{A}}$ in bond-lengths and
$\approx 1^\circ$ in the H-C-H angle with respect to the MP2/6-31G(d)
optimized geometry. The states in Table \ref{tab:formaldehyde} are
ordered according to:
\begin{itemize}
  \item The CIS and CIS-MP2 states are listed in increasing order
    according to the CIS-MP2 excitation energies.
  \item Experimental vertical excitations are listed according to
    the assignment provided in Ref. \onlinecite{hadad1993} with
    respect to CIS results.
  \item MR-CISD results are listed in increasing order, trying to
    match the assignments provided in Ref. \onlinecite{pitarch2003}
    with those in Ref. \onlinecite{hadad1993}.
  \item The $C_{2v}$S-UHF excitation energies are listed in increasing
    order as we have not tried to make a formal assignment of the
    excitation energies. Such an assignment would only provide a guide
    for interpretation as states of the same symmetry always possess
    mixed character.
\end{itemize}

\begingroup
\squeezetable
\begin{table}[!htbp]
  \caption{Vertical excitation energies (in eV) of formaldehyde as
    predicted by the excited $C_{2v}$S-UHF, CIS, CIS-MP2, and
    (SC)$^2$-MR-CISD methods compared to the available experimental
    data. The states are listed according to their spin and
    spatial-symmetry labels. (See text for a description of the
    ordering used in listing the states.)}
  \label{tab:formaldehyde}
  \begin{ruledtabular}
  \begin{tabular}{l @{\hspace{0.3cm}} *{5}{D{.}{.}{6.2}} c @{\hspace{0.3cm}} *{5}{D{.}{.}{6.2}}}
      & \multicolumn{5}{c}{singlets} && \multicolumn{5}{c}{triplets} \\ %
         \cline{2-6} \cline{8-12}
      label
      & \multicolumn{1}{r}{EPHF\footnotemark[1]}
      & \multicolumn{1}{r}{CIS\footnotemark[2]}
      & \multicolumn{1}{r}{CIS-MP2\footnotemark[2]}
      & \multicolumn{1}{r}{MRCI\footnotemark[3]}
      & \multicolumn{1}{r}{expt\footnotemark[4]} &
      & \multicolumn{1}{r}{EPHF\footnotemark[1]}
      & \multicolumn{1}{r}{CIS\footnotemark[5]}
      & \multicolumn{1}{r}{CIS-MP2\footnotemark[2]}
      & \multicolumn{1}{r}{MRCI\footnotemark[3]}
      & \multicolumn{1}{r}{expt\footnotemark[4]} \\\hline
$A_1$ &  9.01 &  9.66 &  8.47 &  8.27 &  8.14 &&  6.08 &  4.65 &  6.72 &  6.05 &  5.86 \\
      &  9.25 & 10.88 &  8.75 &  9.31 &       &&  8.67 &  9.31 &  7.78 &  8.15 &  7.96 \\
      &  9.66 &  9.45 &  9.19 &  9.68 &       && 10.52 & 10.56 &  9.12 &  9.35 &  9.59 \\
      & 10.37 & 11.24 &  9.20 &       &       && 10.85 & 10.88 &       &  9.64 &       \\
      & 11.90 & 12.09 &  9.99 &       &       && 11.81 & 11.89 &       &       &       \\[8pt]
$A_2$ &  4.10 &  4.48 &  4.58 &  4.04 &  4.07 &&  4.30 &  3.67 &  4.15 &  3.59 &  3.50 \\
      &  8.62 &  9.78 &  7.83 &  8.36 &  8.37 &&  8.94 &  9.72 &  8.16 &  8.41 &       \\
      &  9.91 & 10.92 & 10.08 &  9.34 &  9.22 && 10.22 & 10.20 & 10.52 &  9.37 &       \\
      & 11.82 & 12.06 & 10.13 &       &       && 11.17 & 11.15 &       &       &       \\
      &       & 11.37 & 10.49 &       &       &&       &       &       &       &       \\
      &       & 11.93 & 11.63 &       &       &&       &       &       &       &       \\[8pt]
$B_1$ &  9.53 &  9.66 &  9.97 &  9.33 &       &&  9.07 &  8.37 &  9.18 &  8.52 &       \\
      & 10.84 & 11.05 & 10.84 &       & 10.60 && 10.82 & 10.86 &       &  9.33 &       \\
      & 11.72 & 11.84 & 11.56 &       & 11.70 && 13.75 & 11.56 &       &       &       \\[8pt]
$B_2$ &  7.62 &  8.63 &  6.85 &  7.12 &  7.11 &&  7.76 &  8.28 &  6.97 &  6.98 &  6.83 \\
      &  8.40 &  9.36 &  7.66 &  7.95 &  7.97 &&  8.55 &  9.04 &  7.75 &  7.81 &  7.79 \\
      &  9.49 & 10.61 &  8.46 &  8.96 &  8.88 &&  9.68 & 10.33 &  8.67 &  8.83 &       \\
      &  9.77 & 10.86 &  8.94 &  9.18 &       &&  9.97 & 10.69 &       &  9.16 &       \\
      &  9.83 & 10.98 &  8.96 &  9.27 &       && 10.04 & 10.80 &       &       &       \\
      & 10.19 & 11.17 &  9.19 &       &       && 10.30 & 11.07 &       &       &       \\
      & 11.11 &       &       &       &       && 11.38 &       &       &       &       \\
  \end{tabular}
  \end{ruledtabular}
  \footnotetext[1]{$C_{2v}$-SUHF; this work.}
  \footnotetext[2]{From Ref. \onlinecite{hadad1993}.}
  \footnotetext[3]{(SC)$^2$-MR-CISD results from
    Ref. \onlinecite{pitarch2003}.}
  \footnotetext[4]{Experimental excitation energies from
    Refs. \onlinecite{robin,taylor1982,hadad1993}.}
  \footnotetext[5]{CIS; this work.}
\end{table}
\endgroup

The results shown in Table \ref{tab:formaldehyde} show a surprisingly
good qualitative agreement between our $C_{2v}$S-UHF calculations and
the MR-CISD and experimental vertical excitation energies. In
particular, the $C_{2v}$S-UHF results significantly improve over CIS
for most of the excitations listed in the table. Larger deviations are
observed for states with significant Rydberg character, suggesting
that the basis set used is still not sufficient to converge such
excitation energies. The observed agreement between $C_{2v}$S-UHF
excitation energies with best theoretical estimates is encouraging as
it showcases the ability of the simple, mean-field excited
symmetry-projected HF approach to describe excited states of molecular
systems.

\subsection{$C_{2v}$ insertion pathway for BeH$_2$}

The model $C_{2v}$ insertion pathway of Be into H$_2$ is a known
challenging system. It was originally proposed by Purvis {\em et al.}
\cite{purvis1983} as testing ground for single-reference
coupled-cluster methods. It has recently been used as a benchmarking
case for novel multi-reference based methods (see
Ref. \onlinecite{evangelista2006} and references therein).

We follow Ref. \onlinecite{evangelista2006} in the construction of the
model pathway: with the beryllium atom placed at the origin, the $y$
coordinates (in bohr) of the hydrogen atoms are related to their $x$
coordinates (in bohr) by the equation
\[
  y(x) = \pm (2.54 - 0.46x) \qquad x \in [0,4].
\]
At $x=0$, the geometry described corresponds approximately to the
BeH$_2$ equilibrium geometry, while at $x=4$, the geometry corresponds
to a hydrogen molecule at equilibrium interacting with a Be atom
placed $4$ bohr away. A linear interpolation is used for intermediate
geometries; the model insertion pathway has $C_{2v}$ symmetry. In our
calculations, we use the same small basis set as that used in
Ref. \onlinecite{evangelista2006}, corresponding to the contraction
scheme Be($10s3p/3s2p$) and H($4s/2s$).

The BeH$_2$ model insertion pathway is challenging as the dominant
configurations in the FCI expansion change character. At the
equilibrium BeH$_2$ geometry, the dominant configuration is
\[
  (1\sigma_g)^2 \, (2\sigma_g)^2 \, (1\sigma_u)^2 \qquad \equiv \qquad
  (1a_1)^2 \, (2a_1)^2 \, (1b_2)^2,
\]
where the l.h.s. configuration uses the $D_{\infty h}$ symmetry of the
linear molecule and the r.h.s. is its representation in the $C_{2v}$
subgroup. On the other hand, at dissociation, the dominant
configuration becomes
\[
  (1a_1)^2 \, (2a_1)^2 \, (3a_1)^2.
\]
Note hat the latter corresponds to a double excitation with respect to
the reference determinant near the BeH$_2$ equilibrium. Excited state
methods based on a particle-hole construction out of a
symmetry-adapted reference would therefore fail to provide even a
qualitatively correct profile for the first excited
state. Single-reference coupled-cluster can correctly describe the
ground-state dissociation pathway only when different references are
used in different intervals of $x$ \cite{purvis1983,evangelista2006};
the resulting curve is, nevertheless, discontinuous. We note that
different flavors of multi-reference coupled-cluster can correctly
describe the model insertion pathway \cite{evangelista2006}.

\begin{figure}[!htbp]
  \centering
  \includegraphics[width=0.85\textwidth]{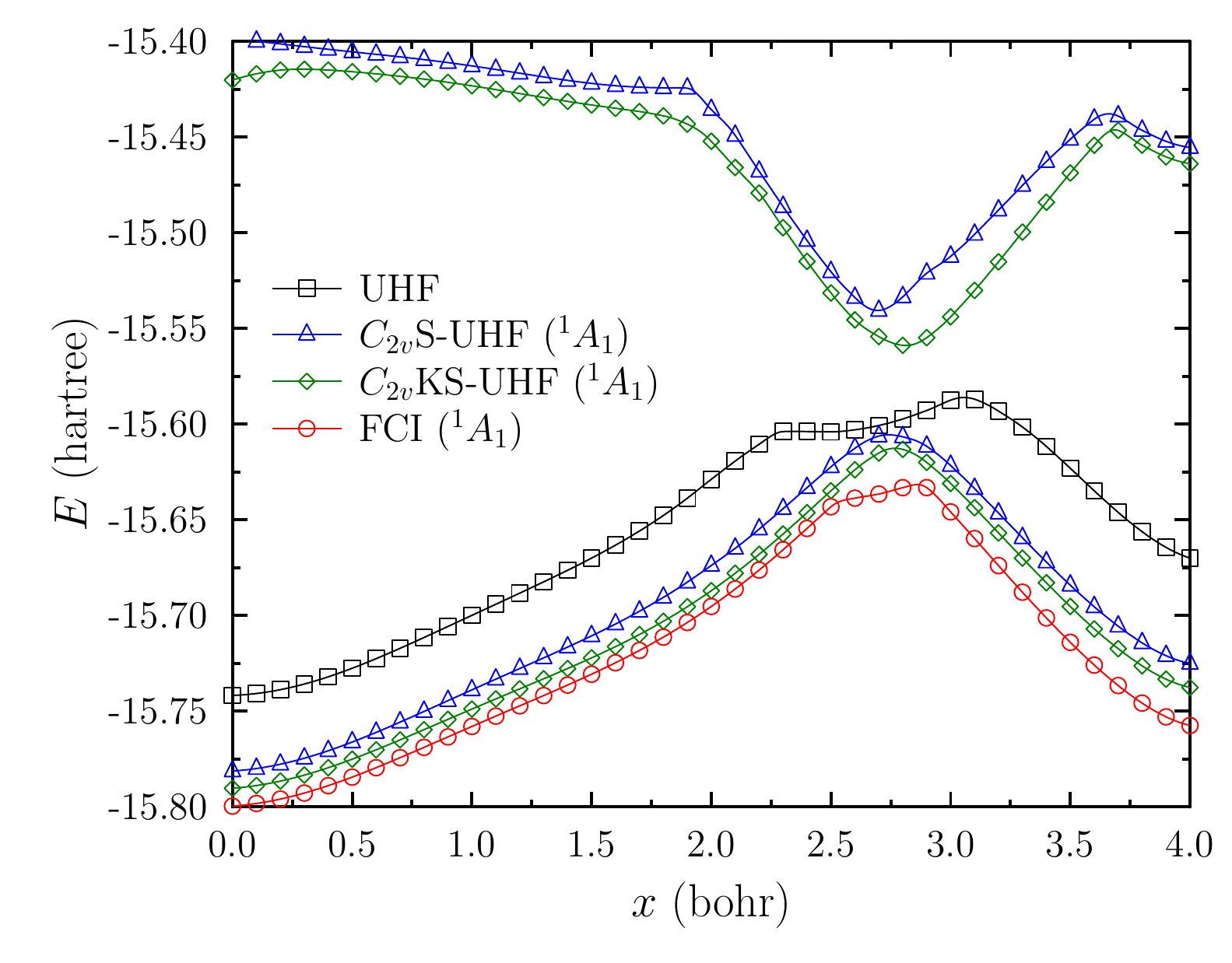}
  \caption{Model $C_{2v}$ insertion pathway of Be into H$_2$ as a
    function of the $x$-coordinate of the hydrogen atoms (Be is placed
    at the origin). The FCI results were obtained from
    Ref. \onlinecite{evangelista2006}. Note that $C_{2v}$S-UHF and
    $C_{2v}$KS-UHF predict smooth curves for the two low-lying $^1
    A_1$ states.}
  \label{fig:beh2}
\end{figure}

We show in Fig. \ref{fig:beh2} the insertion pathways predicted by
UHF, $C_{2v}$S-UHF, and $C_{2v}$KS-UHF, as a function of the $x$
coordinate of the hydrogen atoms. In $C_{2v}$S-UHF, the full-spin
symmetry and the spatial symmetry ($C_{2v}$) are broken and restored
self-consistently; in $C_{2v}$KS-UHF, we additionally break and
restore complex conjugation (denoted by K). We also present the ground
state FCI curve (obtained from Ref. \onlinecite{evangelista2006}) for
comparison purposes.

As Fig. \ref{fig:beh2} shows, both excited symmetry-projected HF
approaches provide smooth curves for both the ground state and the
first excited state. The obtained curves display the interaction
between the lowest $^1 A_1$ states in the model reaction
pathway. Moreover, the profiles are qualitatively similar to the
CASSCF and FCI curves reported in Ref. \onlinecite{purvis1983} for
both states. The $C_{2v}$KS-UHF curve lies very close to the FCI curve
in the interval $x < 2.5$. Near $x \approx 2.7$, where the
multi-reference character is expected to be highest as the different
configurations interact strongly, both $C_{2v}$S-UHF and
$C_{2v}$KS-UHF deviate from the ground state FCI curve. We stress,
nevertheless, that our results can be improved by: a) using more
symmetry-projected configurations for each state, and b) including
more states in the excited symmetry-projected HF approach.

\section{Conclusions}
\label{sec:conclusions}

The spectrum of a molecular system constitutes the fingerprint of its
quantum mechanical character. Characterization of the low-lying
excited states of a system is of paramount importance in order to
understand photochemical and photophysical processes occurring in
nature. Accessing an excited state of the same symmetry as the ground
state has always been challenging for variational strategies. This is
because, if the optimization is carried out using the same formalism
as that used for the ground state, a variational collapse is almost
inevitable. In the formalism discussed in this work we avoided this
collapse by using an ansatz that is explicitly orthogonal to states of
the same symmetry previously obtained. We note that the Gram-Schmidt
orthogonal construction used in this work in terms of
symmetry-projected configurations could be used with other types of
wavefunction.

In a nutshell, our formalism uses chain of variational calculations to
characterize the low-lying excited states of a system with a given set
of quantum numbers in terms of symmetry-projected configurations. The
use of the latter implies that the wavefunctions thus obtained have
well defined symmetries.

We have applied the excited symmetry-projected HF formalism to
describe the dissociation profile of the C$_2$ molecule,
to characterize the low-lying spectrum of formaldehyde, and to explore
a model insertion pathway for BeH$_2$. Several features of the
potential energy curve of the carbon dimer were correctly reproduced;
in particular, the non-Morse shape of the lowest lying $A_1$ state is
obtained after the two symmetry-projected configurations are allowed
to interact. This constitutes the avoided crossing also observed with
other multi-configurational methods such as MRCI or CASPT2. The
low-lying singlet and triplet spectrum of formaldehyde was
characterized and compared with available experimental adiabatic
excitation energies. We have observed a good agreement between our
computed spectrum and the experimental one (all excitation energies
are correct within a $\approx 1$ eV window). This is remarkable given
that each state was essentially described by a single-symmetry
projected configuration, that is, we solved for as many states as the
amount of symmetry-projected configurations we used.

The methodology here considered can be applied to larger systems as it
has mean-field cost. We believe that this method can become a useful
tool for the computational chemist. It may be used to fill the void
between the high-accuracy methods and the large-scale methods, where
the latter typically assume a particle-hole character for the
low-lying excited states.

\section*{Acknowledgments}
\label{sec:acknowledge}

This work was supported by the Department of Energy, Office of Basic
Energy Sciences, Grant No. DE-FG02-09ER16053. G.E.S. is a Welch
Foundation Chair (C-0036). C.A.J.H. acknowledges support from the
Lodieska Stockbridge Vaughn Fellowship.

\appendix

\section{Matrix elements between symmetry-projected configurations}
\label{sec:matrix_elements}

In this appendix, we provide explicit expressions for matrix elements
required in the evaluation of the energy and the energy gradient of
wavefunctions based on symmetry-projected configurations. As shown
below, these can be in turn written in terms of matrix elements
between (non-orthogonal) rotated Slater determinants. L\"owdin
\cite{lowdin1955} first described the evaluation of arbitrary operator
matrix elements between non-orthogonal $N$-electron Slater
determinants. An extended Wick's theorem can be used to evaluate such
matrix elements as shown by, {\em e.g.}, Blaizot and Ripka
\cite{blaizot_ripka}.

We begin by describing the notation used. We work with a set of $M$
elementary fermion creation $\{ c^\dagger \}$ and annihilation $\{ c
\}$ operators satisfying the standard anti-commutation rules
\begin{subequations}
  \begin{align}
    \Big[ c_j, c_k \Big]_+ &= \, 0, \\
    \Big[ c_j^\dagger, c_k^\dagger \Big]_+ &= \, 0, \\
    \Big[ c_j, c_k^\dagger \Big]_+ &= \, \langle j | k \rangle =
      \delta_{jk}.
  \end{align}
\end{subequations}
Note than an orthonormal basis is used. The transformation from the
non-orthogonal atomic orbital basis to an orthonormal one is
straightforward.

The non-relativistic, Born-Oppenheimer molecular electronic
Hamiltonian $\hat{H}$ is expressed in second quantization as
\begin{equation}
  \hat{H} = \sum_{ik} \langle i | \hat{h} | k \rangle \, c_i^\dagger
    \, c_k + \frac{1}{4} \sum_{ijkl} \langle ij | \hat{v} | kl \rangle
    \, c_i^\dagger \, c_j^\dagger \, c_l \, c_k,
\end{equation}
where $\langle i | \hat{h} | k \rangle$ are one-electron (core
Hamiltonian) integrals and $\langle ij | \hat{v} | kl \rangle$ are
anti-symmetrized two-electron (electron repulsion) integrals in Dirac
notation.

An arbitrary $N$-electron Slater determinant $|\Phi \rangle$ is
constructed as a vacuum to a set of $N$ occupied (hole) HF creation
operators $\{ b_h^\dagger \, | \, h = 1, \ldots, N \}$ and $M-N$
virtual (particle) HF annihilation operators $\{ b_p \, | \, p = N+1,
\ldots, M \}$. It may be represented as
\begin{equation}
  |\Phi \rangle = \prod_{h=1}^N b_h^\dagger |- \rangle,
\end{equation}
where $|- \rangle$ is the bare fermion vacuum (annihilated by $\{ c
\}$). The HF operators are given as linear combinations of the
elementary fermion ones, that is,
\begin{equation}
  b_k^\dagger = \sum_j D_{jk}^\ast \, c_j^\dagger,
\end{equation}
where $D$ is the matrix of molecular orbital coefficients. (Note that
our choice for the matrix of orbital coefficients is the complex
conjugate of the standard one.) As usual, the first $N$ columns in $D$
are used for the occupied molecular orbitals, while the remaining
columns describe the virtual orbitals. The above transformation is
canonical (it preserves fermion anti-commutation rules) if the matrix
$D$ is unitary: $D \, D^\dagger = D^\dagger \, D = \mathbf{1}$.

In order to provide explicit expressions for the matrix elements, we
introduce a generic form of the projection operator $\hat{P}^j_{mk}$
(for general non-Abelian groups) given by
\begin{equation}
  \hat{P}^j_{mk} = \frac{1}{V} \int_V d\vartheta \, w^j_{mk}
    (\vartheta) \, \hat{R} (\vartheta).
  \label{eq:gen_proj}
\end{equation}
A state transforming as the $m$-th row of the $j$-th irreducible
representation is recovered upon the action of the above projection
operator on an arbitrary state. Here, $\vartheta$ labels the elements
of the symmetry group; for discrete groups, the integration should be
understood as a summation. In addition, $V$ is the volume of
integration, $w^j_{mk} (\vartheta)$ is an integration weight
(character) associated with the symmetries of the state to be
recovered, and $\hat{R} (\vartheta)$ is a rotation operator.

For all the cases considered in this work, $\hat{R} (\vartheta)$ is a
single-particle rotation operator that transforms the HF operators
according to
\begin{align}
  b_k^\dagger (\vartheta) &\equiv \, \hat{R} (\vartheta) \,
    b_k^\dagger \, \hat{R}^{-1} (\vartheta) \nonumber \\
  &= \, \sum_j D_{jk}^\ast \, \hat{R} (\vartheta) \, c_j^\dagger \,
    \hat{R}^{-1} (\vartheta) = \sum_{ji} D_{jk}^\ast \, R_{ij}
    (\vartheta) \, c_i^\dagger,
\end{align}
where $R_{ij} (\vartheta) = \langle i | \hat{R} (\vartheta) | j
\rangle$ is an element of the matrix representation of the rotation
operator in the single-particle basis.

Using Eq. \ref{eq:gen_proj}, overlap and Hamiltonian matrix elements
between symmetry-projected configurations are expressed in terms of
norm and Hamiltonian overlaps between rotated determinants $\hat{R}
(\vartheta) |\Phi \rangle$ as
\begin{subequations}
  \begin{align}
    \langle \Phi^r | \hat{P}^j_{kk'} | \Phi^s \rangle
    &= \, \frac{1}{V} \int_V d\vartheta \, w^j_{kk'} (\vartheta) \,
      n^{rs} (\vartheta), \\[4pt]
    \langle \Phi^r | \hat{H} \, \hat{P}^j_{kk'} | \Phi^s \rangle
    &= \, \frac{1}{V} \int_V d\vartheta \, w^j_{kk'} (\vartheta) \,
      n^{rs} (\vartheta) \, h^{rs} (\vartheta),
  \end{align}
\end{subequations}
where
\begin{subequations}
  \begin{align}
    n^{rs} (\vartheta) &\equiv \,
    \langle \Phi^r | \hat{R} (\vartheta) | \Phi^s \rangle,
    \label{eq:def_nrs} \\[4pt]
    h^{rs} (\vartheta) &\equiv \,
    \frac{\langle \Phi^r | \hat{H} \, \hat{R} (\vartheta) | \Phi^s
      \rangle}{\langle \Phi^r | \hat{R} (\vartheta) | \Phi^s \rangle}.
    \label{eq:def_hrs}
  \end{align}
\end{subequations}

The norm overlaps of Eq. \ref{eq:def_nrs} can be evaluated by applying
Wick's theorem on the bare fermion vacuum. This leads to
\begin{align}
  n^{rs} (\vartheta) &= \, \mathrm{det}_N \, X^{rs} (\vartheta), \\
  X^{rs} (\vartheta) &= \, D^{r\trans} \, R(\vartheta) \, D^{s\ast}.
  \label{eq:def_xrs}
\end{align}
Here, $D^k$ is the (rectangular) matrix of occupied orbital
coefficients associated with the determinant $|\Phi^k \rangle$ and
$R(\vartheta)$ is the matrix representation of the rotation operator
in the single-particle basis. The notation $\mathrm{det}_N$ is used to
emphasize that the determinant should be evaluated over the $N \times
N$ block of $X^{rs} (\vartheta)$ defined by the occupied states in
$|\Phi^r \rangle$ and $|\Phi^s \rangle$.

The Hamiltonian overlaps of Eq. \ref{eq:def_hrs} can be evaluated by
using an extended Wick's theorem \cite{blaizot_ripka} when $|\Phi^r
\rangle$ and $|\Phi^s \rangle$ are not orthogonal. They are given by
\begin{align}
  h^{rs} (\vartheta) &= \, \sum_{ik} \left[ \langle i | \hat{h} | k
    \rangle + \frac{1}{2} \, \Gamma^{rs}_{ik} (\vartheta) \right]
    \rho^{rs}_{ki} (\vartheta), \\
  \Gamma^{rs}_{ik} (\vartheta) &= \, \sum_{jl} \langle ij | \hat{v} |
    kl \rangle \, \rho^{rs}_{lj} (\vartheta).
\end{align}
The Hamiltonian overlaps are expressed in terms of the transition
density matrix $\rho^{rs} (\vartheta)$, with elements defined by
\begin{equation}
  \rho^{rs}_{ki} (\vartheta) \equiv \frac{\langle \Phi^r | c_i^\dagger
    \, c_k \, \hat{R} (\vartheta) | \Phi^s \rangle}{\langle \Phi^r |
    \hat{R} (\vartheta) | \Phi^s \rangle}.
  \label{eq:def_rhors}
\end{equation}
The transition density matrix of Eq. \ref{eq:def_rhors} is built
according to \cite{lowdin1955}
\begin{equation}
  \rho^{rs} (\vartheta) = R(\vartheta) \, D^{s\ast} \, \left[ X^{rs}
    (\vartheta) \right]^{-1} \, D^{r\trans}.
\end{equation}
Here, the inverse of $X^{rs} (\vartheta)$ (defined in
Eq. \ref{eq:def_xrs}) should be evaluated over the $N \times N$ block
of occupied states in both determinants. Accordingly, only the
occupied orbitals in $D^r$ and $D^s$ should be used in computing the
matrix product above.

Matrix elements appearing in contributions to the energy gradient can
also be expressed in terms of overlaps between rotated determinants:
\begin{subequations}
  \begin{align}
    \langle \Phi^r | b_h^{r\dagger} \, b_p^r \, \hat{P}^j_{kk'} |
      \Phi^s \rangle
    &= \, \frac{1}{V} \int_V d\vartheta \, w^j_{kk'} (\vartheta) \,
      n^{rs} (\vartheta) \, N^{rs}_{ph} (\vartheta), \\[4pt]
    \langle \Phi^r | b_h^{r\dagger} \, b_p^r \, \hat{H} \,
      \hat{P}^j_{kk'} | \Phi^s \rangle
    &= \, \frac{1}{V} \int_V d\vartheta \, w^j_{kk'} (\vartheta) \,
      n^{rs} (\vartheta) \, H^{rs}_{ph} (\vartheta),
  \end{align}
\end{subequations}
where we have appended a superscript to the HF operators to label the
determinant to which they are associated. Here,
\begin{subequations}
  \label{eq:grad_rs}
  \begin{align}
    N^{rs}_{ph} (\vartheta) &\equiv \,
    \frac{\langle \Phi^r | b_h^{r\dagger} \, b_p^r \, \hat{R}
      (\vartheta) | \Phi^s \rangle}{\langle \Phi^r | \hat{R}
      (\vartheta) | \Phi^s \rangle},
    \\[4pt]
    H^{rs}_{ph} (\vartheta) &\equiv \,
    \frac{\langle \Phi^r | b_h^{r\dagger} \, b_p^r \, \hat{H} \,
      \hat{R} (\vartheta) | \Phi^s \rangle}{\langle \Phi^r | \hat{R}
      (\vartheta) | \Phi^s \rangle}.
  \end{align}
\end{subequations}

The matrix elements of Eq. \ref{eq:grad_rs} can be evaluated using an
extended Wick's theorem when $|\Phi^r \rangle$ and $|\Phi^s \rangle$
are not orthogonal. They are given by
\begin{subequations}
  \begin{align}
    N^{rs}_{ph} (\vartheta) &= \,
    \left[ D^{r\trans} \, \rho^{rs} (\vartheta) \, D^{r\ast}
      \right]_{ph},
    \\[4pt]
    H^{rs}_{ph} (\vartheta) &= \,
    h^{rs} (\vartheta) \left[ D^{r\trans} \, \rho^{rs} (\vartheta) \,
      D^{r\ast} \right]_{ph}
    \nonumber \\
    &+ \, \left[ D^{r\trans} \, \left( \mathbf{1} -
      \rho^{rs} (\vartheta) \right) \, f^{rs} (\vartheta) \,
      \rho^{rs} (\vartheta) \, D^{r\ast} \right]_{ph},
  \end{align}
\end{subequations}
where we have set $f^{rs}_{ik} (\vartheta) = \langle i | \hat{h} | k
\rangle + \Gamma^{rs}_{ik} (\vartheta)$.

%

\end{document}